\documentclass{PoS}

\def\Vec#1{\mbox{\boldmath $#1$}}
\def\beq{\begin{equation}}
\def\eeq{\end{equation}}
\def\beqy{\begin{eqnarray}}
\def\eeqy{\end{eqnarray}}

\title{Initial state anisotropies in ultrarelativistic 
  heavy-ion collisions from the Monte Carlo Glauber model}

\ShortTitle{Initial state anisotropies uncertainties in MCG}

\author{\speaker{M. Alvioli}\\
  ECT*, European Centre for Theoretical Studies in Nuclear Physics and Related Areas\\
  Strada delle Tabarelle 286, I-38123 Villazzano (TN) Italy\\
  E-mail: \email{alvioli@pg.infn.it}}

\author{H. Holopainen\\
  Frankfurt Institute for Advanced Studies, Ruth-Moufang-Strasse 1, D-60438 Frankfurt am Main, Germany\\
  E-mail: \email{holopainen@fias.uni-frankfurt.de}}

\author{K. J. Eskola\\
  Department of Physics, P.O.Box 35, FI-40014 University of Jyv\"askyl\"a, Finland\\
  E-mail: \email{kari.eskola@phys.jyu.fi}}

\author{M. Strikman\\
  The Pennsylvania State University, 104 Davey Lab, University Park, Pennsylvania 16803, USA\\
  E-mail: \email{strikman@phys.psu.edu}}

\abstract{In hydrodynamical modeling of heavy-ion collisions the initial state spatial 
  anisotropies translate into momentum anisotropies of the final state particle distributions. 
  Thus, understanding the origin of the initial anisotropies and quantifying their uncertainties 
  is important for the extraction of specific QCD matter properties, such as viscosity, from the 
  experimental data. In this work we study the wounded nucleon approach in the Monte Carlo Glauber 
  model framework, focusing especially on the uncertainties which arise from the modeling of the 
  nucleon-nucleon interactions between the colliding nucleon pairs and nucleon-nucleon correlations 
  inside the colliding nuclei. We compare the black disk model and a probabilistic profile function 
  approach for the inelastic nucleon-nucleon interactions, and study the effects of initial state 
  correlations using state-of-the-art modeling of these.}

\FullConference{Sixth International Conference on Quarks and Nuclear Physics,\\
		April 16-20, 2012\\
		Ecole Polytechnique, Palaiseau, Paris}

\begin{document}
\section{Introduction}

In ultrarelativistic heavy-ion collisions performed at the Relativistic Heavy Ion
Collider (RHIC) and Large Hadron Collider (LHC) significant azimuthal momentum
distribution anisotropies have been measured \cite{SWcern,ALICE:2011ab}. These
anisotropies can be explained with relativistic hydrodynamics: the initially produced
QCD-matter contains spatial anisotropies and during the hydrodynamical evolution these
anisotropies are transferred to the momentum distributions of final state particles.

Simulations with viscous hydrodynamics have shown that the shear viscosity of the
QCD-matter produced in ultrarelativistic heavy-ion collisions can be
estimated from the final state momentum anisotropies. Since the origin of these
anisotropies is in the initial state, uncertainties related to the initial state must
be charted before reliable estimates for the viscosity can be made. In this
work \cite{Alvioli:2011sk} we consider two sources of uncertainties related to the Monte
Carlo Glauber (MCG) model which is often used to initialise the hydrodynamical
simulations (see \textit{e.g.} \cite{Holopainen:2010gz}, \cite{Qin:2010pf}, \cite{Rybczynski:2011wv}).

The Glauber model \cite{glauber} is usually a key element in computing the initial states for  
hydrodynamical modeling of ultrarelativistic heavy-ion collisions. Some years back, most 
hydrodynamical calculations assumed smooth initial states where the (energy or entropy) densities 
were assumed to scale with the density of binary collisions or wounded nucleons computed from the 
optical Glauber model; see, e.g. \cite{Kolb:2001qz}. Now that the importance of the initial density 
fluctuations has been realized, Monte Carlo Glauber (MCG) modeling has become more frequently used. 
So far the black disk (hard-sphere) modeling of the nucleon-nucleon ($NN$) interactions has been the 
standard choice \cite{Holopainen:2010gz,Alver:2008aq,Alver:2008zza,Hirano:2009ah} in these studies, 
although also more involved probabilistic ways to model the $NN$ interactions have been known for a 
long time \cite{glauber,Pi:1992ug,Wang:1990qp,Glauber:1970jm}.
In Ref.~\cite{Alvioli:2011sk} we studied \textit{black disk} and \textit{profile function} models.

In the MCG modeling one needs to know the positions of the initial state nucleon configurations. 
In most cases the nucleon positions inside the colliding nuclei are just sampled using the 
Woods-Saxon potential, neglecting nucleon-nucleon correlations \cite{Subedi:2008zz}.
However, there exist calculations which show that high-momentum components of the nuclear wave 
function are generated by the two-body $NN$ correlations \cite{Alvioli:2005cz,Alvioli:2007zz}. 

In Ref. \cite{Alvioli:2011sk} we studied for Au-Au collision at RHIC center of mass energy 
$\sqrt{s_{NN}}$=200 GeV, two different uncertainties in computing the initial state asymmetries 
from the MCG model: one related to the modeling of the inelastic $NN$ collisions between nucleons 
from different nuclei, and one related to the $NN$ correlations in the nucleon configurations in 
each of the colliding nuclei. In this contribution we extend the results from Ref. \cite{Alvioli:2011sk} 
about the anisotropy moments $\epsilon_n$ with $n=1,2,3,4,5$: dipole asymmetry, eccentricity, triangularity 
\cite{Alvioli:2011sk}, quadangular and  pentagonal asymmetries. Results will be shown for Au+Au collisions 
at RHIC with an inelastic nucleon-nucleon cross section $\sigma_{NN}=42$~mb.

\section{MCG framework: nucleon configurations}
\label{sec: nucl conf}

The initial state of a nucleus in the MCG calculations is usually 
taken as a collection of particles distributed according to a probability distribution 
given by the corresponding (Woods-Saxon) number density distribution measured in electron scattering
experiments. Given the complexity of the nuclear many-body problem, the effects of spatial, spin and
isospin dependent correlations among the nucleons are usually overlooked and the nucleons are 
positioned randomly for each of the simulated events.
Recently, in Ref.~\cite{Alvioli:2009ab} it was shown how such an approach can be modified by including 
initial states, which are prepared in advance, in the commonly used computer codes. Also, a method to 
produce such configurations was introduced. The method is based on the notion of a nuclear wave function 
$\psi$, which contains the nucleonic degrees of freedom and which is used to iteratively modify the positions 
of randomly distributed nucleons using the Metropolis method so that the final positions correspond to the 
probability density given by $|\psi|^2$. The method is constructed to reproduce the same nucleon number-density 
distribution as the usual one and, in addition, to reproduce the basic features of the two-nucleon density in 
the presence of the $NN$ correlations. The model wave function is taken in the form
\beq
\psi(\Vec{x}_1,...,\Vec{x}_A)=\prod^A_{i<j}\hat{f}_{ij}\,\phi(\Vec{x}_1,...,\Vec{x}_A)\,,
\eeq
where $\phi$ is the uncorrelated wave function and $\hat{f}_{ij}$ are correlation
operators; here, $\Vec{x}_i$ denotes the position, spin and isospin projection of the
$i$-th nucleon. The correlation operator contains a detailed spin-isospin dependence.
In the most general case, this dependence includes a number of channels that are
the same as the one appearing in modern nucleon-nucleon potentials used to successfully 
describe a variety of properties of light and medium-heavy nuclei within different 
\textit{ab initio} approaches. The major merit of this approach is that the two-body 
densities resulting from configurations obtained using these correlations are clearly 
more realistic than the completely uncorrelated ones \cite{Alvioli:2010yk}.
In this paper we have used configurations generated by two-body correlations, including the 
tensor operator, and three-body clusters surrounding each of the nucleons, induced by full 
correlations. The effects of genuine three-body correlations have been discussed in Ref. 
\cite{Alvioli:2011sk}.

\subsection{MCG framework: modeling the inelastic interactions}
\label{sec: interaction}

In each simulated event, given the impact parameter of the $A+A$ collision, the
nucleon-nucleon interactions must be modeled.
We work in the Glauber model framework \cite{glauber}, neglecting the effects of inelastic 
diffraction that lead to fluctuations of the strength of the $NN$ interactions
\cite{Blaettel:1993ah,Baym:1995cz}. To generate the inelastic $NN$ collisions of interest here, 
we use the following two different approximations for establishing whether a collision between 
the nucleons $i$ and $j$ from different nuclei takes place: 

\noindent
\textbullet~
\textit{Black disk} approximation, used recently, e.g., in Ref. \cite{Holopainen:2010gz}, where 
one assumes the two nucleons to interact inelastically with a probability one if their transverse 
separation $b_{ij}$ is within a radius defined by the inelastic $NN$ cross section $\sigma^{in}_{NN}$,
\beq
\label{bdisk}
\Vec{b}^2_{ij}\le \frac{\sigma^{in}_{NN}}{\pi}\,;
\eeq

\noindent
\textbullet~
\textit{Profile function approach}, where the probability of an inelastic interaction between the 
nucleons $i$ and $j$ is given by
\beq
\label{intprob}
P(\Vec{b}_{ij})\,=\,1\,-\,\left|1-\Gamma(\Vec{b}_{ij})\right|^2\,,
\eeq
and where the profile function $\Gamma$ is expressed in terms of the total and elastic $NN$ cross sections 
as follows:
\beq
\label{gamma}
\Gamma(\Vec{b}_{ij})\,=\,\frac{\sigma^{tot}_{NN}}{4 \pi B}\,e^{-b^2_{ij}/(2\,B)}\,,
\eeq
with $B=(\sigma^{tot}_{NN})^2/(16\pi\sigma^{el}_{NN})$.

The probability distribution $P(\Vec{b}_{ij})$ can be derived in the Born approximation of the 
potential-scattering formalism \cite{glauber,Blaettel:1993ah,Alvioli:2008rw}; details are given 
in Ref. \cite{Alvioli:2011sk}. The nucleon-nucleon elastic, total and inelastic cross sections 
are given in this formalism by:
\beqy
\label{ssel}
\sigma^{el}_{NN}&=&\frac{\left(\sigma^{tot}_{NN}\right)^2}{16 \pi B}\,,\\
\label{stot}
\sigma^{tot}_{NN}&=&\sigma^{el}_{NN}\,+\,\sigma^{in}_{NN}\,,\\
\sigma^{in}_{NN}&=&\int d^2\Vec{b}_{ij}\,\left(1\,-\,\left|1\,-\,\Gamma(\Vec{b}_{ij})\right|^2\right)\,.
\eeqy

After the nucleon-nucleon interactions have been determined, we can calculate the initial 
asymmetries from the positions of the nucleons which had experienced at least one collision. 
These are called wounded (or participant) nucleons.

\subsection{Spatial asymmetries and their fluctuations}
\label{sec: anisotropy}

We compute the spatial asymmetries from the wounded nucleon positions which are obtained
from the MCG model as follows:
\begin{equation}
  \label{epsn}
  \epsilon_n =  -\frac{ \langle w(r) \cos (n(\phi - \psi_n) \rangle }{ \langle w(r) \rangle },
\end{equation}
where $w(r)$ is a weight and we choose $w(r) = r^2$ for $n=1,2,3,4,5$, following 
Ref. \cite{Bhalerao:2011bp} (at variance with Ref. \cite{Alvioli:2011sk} where we used 
$w(r) = r^3$ for $n=1$ \cite{Teaney:2010vd} and $w(r) = r^n$ for $n=2,3$).

The orientation angle $\psi_n$ is determined as
\begin{equation}
  \psi_n = \frac{1}{n} \arctan 
\frac{\langle w(r) \sin (n\phi) \rangle}{\langle w(r) \cos (n\phi) \rangle} + \frac{\pi}{n},
\end{equation}
where the arctan must be placed in the correct quadrant. Since the experimental methods typically
measure the root mean square (rms) of the flow coefficients, we also present the rms values of
initial state anisotropies $\sqrt{\langle \epsilon^2_n\rangle}$ \cite{Bhalerao:2011bp}.
 
\begin{figure}[!htp]
  \centerline{
  \includegraphics[width=8.5cm]{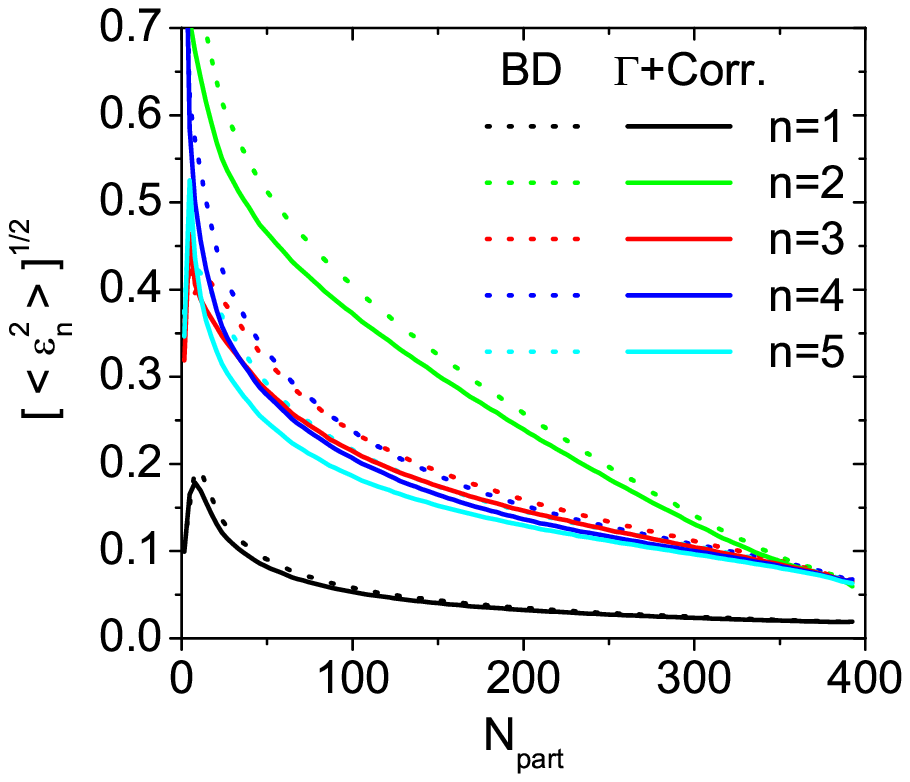}\hspace{-1.0cm}
  \includegraphics[width=8.5cm]{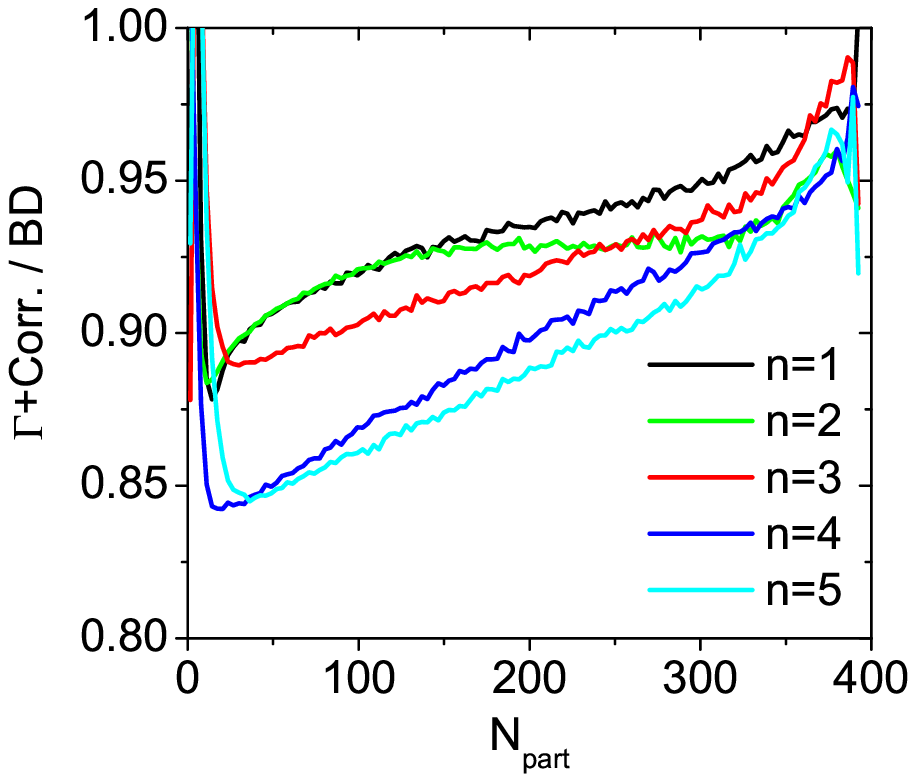}}
  \caption{Left panel: the root mean square of the initial anisotropies,
    $\sqrt{\langle\epsilon^2_n\rangle}$, defined in Eq. (\protect\ref{epsn}),
    as a function of the number of participants. Results are relative to 
    the black disk, uncorrelated approximation (dotted lines) 
    and probabilistic interaction plus correlated configurations (solid lines).
    Right panel: the ratio of the $\Gamma$+correlations over black disk, uncorrelated
    results shown in the left panel.}
  \label{Fig1}
\end{figure}

\section{Results and discussion}
\label{sec: results}

In Ref. \cite{Alvioli:2011sk} we charted some of the uncertainties in the computation of 
the initial state ani\-so\-tropies from the Monte Carlo Glauber model.
A summary of the results is plotted in Fig.~\ref{Fig1}, now using rms values. Also
harmonics $n=4,5$ \cite{Borghini:2005kd,Ollitrault:2009ie}, which were not shown in the original 
article, are shown here. We used two different ways 
of modeling the inelastic interactions between the colliding nucleons. The difference between 
these two cases gives us an estimate about the uncertainties related to this part of the model: 
in central collisions the details of the interaction model play a minor role, but in the peripheral 
collisions such details can cause uncertainties up to 10\% in the first three harmonics $\epsilon_1$, 
$\epsilon_2$ and $\epsilon_3$. The situation is similar for the rms of $\epsilon_n$, as shown
in Fig. \ref{Fig1}, where $n=1,2,3,4,5$ are considered with the choice of the weight function, 
$w(r)=r^2$; the effect is at about 10\% over the $N_{part}$ range and it is mostly due to the
probabilistic profile function approach. We also checked in Ref. \cite{Alvioli:2011sk} that with these 
two interaction models the difference in the number of wounded nucleons and binary collisions remains 
small in central collisions, but at impact parameters 10-15 fm the difference can be around 10\%.

We also presented a study of the effects of $NN$ correlations with an update of correlated configurations 
and extended discussion as compared with the previous published papers on this subject. We confirmed that 
the inclusion of centrally correlated nucleon configurations produce the effects to eccentricity and its 
relative variance as was claimed by Ref.~\cite{Broniowski:2010jd}. As a new result, we observed that the 
inclusion of realistically correlated configurations (two-body full correlations, three-body chains) seems 
to essentially cancel this effect and bring the results back close to the no correlations case. The effect 
is similar for dipole asymmetry and triangularity as for eccentricity. However, we also showed that there 
are still uncertainties caused by the truncation done in the nucleon configuration calculation with full 
correlations and we expect three-body correlations to play a role.

In this Proceedings contributions, we added the study of the root mean square of initial state
ani\-so\-tropies for $n=1,2,3,4,5$ and presented results in Fig.~\ref{Fig1}, with similar sensitivity to
the two sou\-rces of uncertainties studied, which was found to be maximally of the order of 10\%. 
Now that, thanks to the recent developments in event-by-event hydrodynamics and high-precision data,
more precise comparisons of flow coefficients between the data and the theory are becoming possible, 
it is important to quantify all the relevant uncertainties to this precision, so that the QCD matter 
pro\-pe\-rties could eventually be determined from the measured particle spectra and their azimuthal asymmetries.


\end{document}